\documentclass[a4paper,11pt,fleqn]{paper}
\usepackage{hyperref}
\usepackage{epsfig}
\usepackage{exscale}
\usepackage{latexsym}
\usepackage{times}
\usepackage{rotating}

\renewcommand{\thefootnote}{\fnsymbol{footnote}}

\newcommand{\curl}{\mbox{curl}\,}
\newcommand{\ie}{\mathrm{i}\hspace{0.2mm}}
\newcommand{\dn}[1]{_{\protect\mbox{\protect\scriptsize{#1}}}}

\newcommand{\sv}[1]{\tilde{\mbox{\boldmath$\displaystyle{#1}$}}}
 
\newcommand{\nv}[1]{\mbox{\boldmath$\displaystyle{#1}$}}

\newcommand{\mum}{\,\mu\mbox{m}}
\newcommand{\nm}{\,\mbox{nm}}

\newcommand{\PCc}{\mbox{\rm c\hspace{0.2mm}}}

\newcommand{\refeq}[1]{Eq.~(\protect\ref{#1eq})}
\newcommand{\refeqn}[1]{(\protect\ref{#1eq})}

\newcommand{\refmeq}[1]{Eqs.~(\protect\ref{#1eq})}
\newcommand{\labeq}[1]{\label{#1eq}}
\newcommand{\refsec}[1]{Section~\protect\ref{sec#1}}
\newcommand{\refmsec}[1]{Sections~\protect\ref{sec#1}}
\newcommand{\refsecn}[1]{\protect\ref{sec#1}}

\newcommand{\reftwosecs}[2]{Sections~\protect\ref{sec#1}, \protect\ref{sec#2}}
\newcommand{\labsec}[1]{\label{sec#1}}

\newcommand{\labsubsec}[1]{\label{subsec#1}}

\newcommand{\reffig}[1]{Figure~\protect\ref{#1fig}}

\newcommand{\refmfig}[1]{Figures~\protect\ref{#1fig}}
\newcommand{\reffign}[1]{\protect\ref{#1fig}}
\newcommand{\labfig}[1]{\label{#1fig}}
\newcommand{\reftab}[1]{Table~\protect\ref{#1tab}}
\newcommand{\labtab}[1]{\label{#1tab}}

\newcommand{\beq}[1]{\begin{equation}\labeq{#1}\vspace{-1.5mm}}
\newcommand{\eeq}{\vspace{-1.5mm}\end{equation}}
\newcommand{\beqnn}{\begin{displaymath}}
\newcommand{\eeqnn}{\end{displaymath}}
\newcommand{\bea}{\begin{eqnarray}}
\newcommand{\eea}{\end{eqnarray}}

\newcommand{\FIGh}[3]{
\begin{figure}[!htb]
\vspace{2.0mm}
\begin{center}
\epsfig{file=#1.eps,width=#2,clip}
\end{center}
\vspace{-0.7cm}
\caption{#3}
\labfig{#1}
\end{figure}
}

\newcommand{\FIGp}[3]{
\begin{figure}[p]
\vspace{2.0mm}
\begin{center}
\epsfig{file=#1.eps,width=#2,clip}
\end{center}
\vspace{-0.7cm}
\caption{#3}
\labfig{#1}
\end{figure}
}

\newlength{\lefthalf}
\newlength{\righthalf}

\newcommand{\FIGv}[3]{
\setlength{\lefthalf}{0.0cm}
\setlength{\righthalf}{\textwidth}
\addtolength{\lefthalf}{#2}
\addtolength{\lefthalf}{0.2cm}
\addtolength{\righthalf}{-\lefthalf}
\addtolength{\righthalf}{-0.4cm}
\begin{figure}[!htb]
\vspace{2.0mm}
\begin{minipage}[b]{\lefthalf}
\epsfig{file=#1.eps,width=#2,clip}
\end{minipage}
~\hfill~
\begin{minipage}[b]{\righthalf}
\vspace{-0.6cm}
\caption{#3}
\labfig{#1}
\end{minipage}
\end{figure}
}

\newcommand{\TABh}[3]{
\begin{table}[!htb]
\vspace{2.0mm}
\begin{center}
\begin{small}
#2
\end{small}
\end{center}
\vspace{-0.6cm}
\caption{#3}
\labtab{#1}
\end{table}
}

\begin{document}
\begin{sloppypar}

$\mbox{~}$\hfill\begin{rotate}{-90}\sf\footnotesize Manuscript submitted to the IEEE/OSA Journal of Lightwave Technology
\end{rotate}
\vspace{-1cm}

\begin{center}
{\Large\bf 
Full resonant transmission of semi-guided planar waves \\[0.2cm]
through slab waveguide steps at oblique incidence
}
\end{center}

\begin{center}
{\sl
\href{http://www.computational-photonics.eu/m_hammer.html}{Manfred Hammer}\footnote{
\small \sl 
University of Paderborn, FG Theoretical Electrical Engineering \hfill Warburger Str.\ 100, 33098 Paderborn, Germany\\
Phone: ++49(0)5251/60-3560\hfill Fax: ++49(0)5251/60-3524\hfill
E-mail: \href{mailto:manfred.hammer@uni-paderborn.de}{manfred.hammer@uni-paderborn.de}
}, 
\href{http://tet.upb.de/tet/?id=371}{Andre Hildebrandt}, 
\href{http://tet.upb.de/tet/?id=369}{Jens F\"{o}rstner} 
\\[0.1cm]
\href{http://tet.upb.de/}{Theoretical Electrical Engineering}, \href{http://www.uni-paderborn.de/en/}{University of Paderborn}, Germany
}
\end{center}

\vspace{0.1cm}
\begin{center} 
\begin{minipage}{14.1cm}
\rule{14.1cm}{0.1mm}
\begin{small}
\parskip0.1cm
{\bf Abstract:} 
Sheets of slab waveguides with sharp corners  
are investigated. By means of rigorous numerical experiments, we look at 
oblique incidence of semi-guided plane waves. Radiation losses vanish 
beyond a certain critical angle of incidence. One can thus realize lossless 
propagation through 90-degree corner configurations,
where the remaining guided waves are still subject to pronounced reflection
and polarization conversion. A system of two corners can be viewed as a 
structure akin to a Fabry-Perot-interferometer. By adjusting the distance 
between the two partial reflectors, here the 90-degree corners, one identifies 
step-like configurations that transmit the semi-guided plane waves without 
radiation losses, and virtually without reflections. Simulations of 
semi-guided beams with in-plane wide Gaussian profiles show that the effect 
survives in a true 3-D framework.  


\end{small} \\
\rule[1.0ex]{14.1cm}{0.1mm}
\end{minipage}
\end{center}

\renewcommand{\thefootnote}{\arabic{footnote}}
\setcounter{footnote}{0}


\section{Introduction}

In a conventional 2-D setting, any abrupt discontinuities in a high-contrast dielectric optical
slab waveguide typically lead to pronounced radiative losses. This applies also 
to waveguides with sharp corners. Smooth transitions in the form of waveguide 
bends \cite{Mar69,HHS05} help to reduce the losses, but also increase the size of the structures. 
Perhaps this is why the intriguing propagation characteristics \cite{MCK96} of defect waveguides 
in photonic crystals, and of 90$^\circ$ bends made of these, attracted so much attention, despite 
their complexity what concerns simulation, design, and experimental realization. 
With this paper we intend to show that very similar effects can be achieved by much more modest  
means. To this end, corner and step-like structures as illustrated in \reffig{kdk3Dsk} are considered, 
for out-of-plane guided, in-plane unguided plane waves at {\em oblique} angles of incidence.

\FIGv{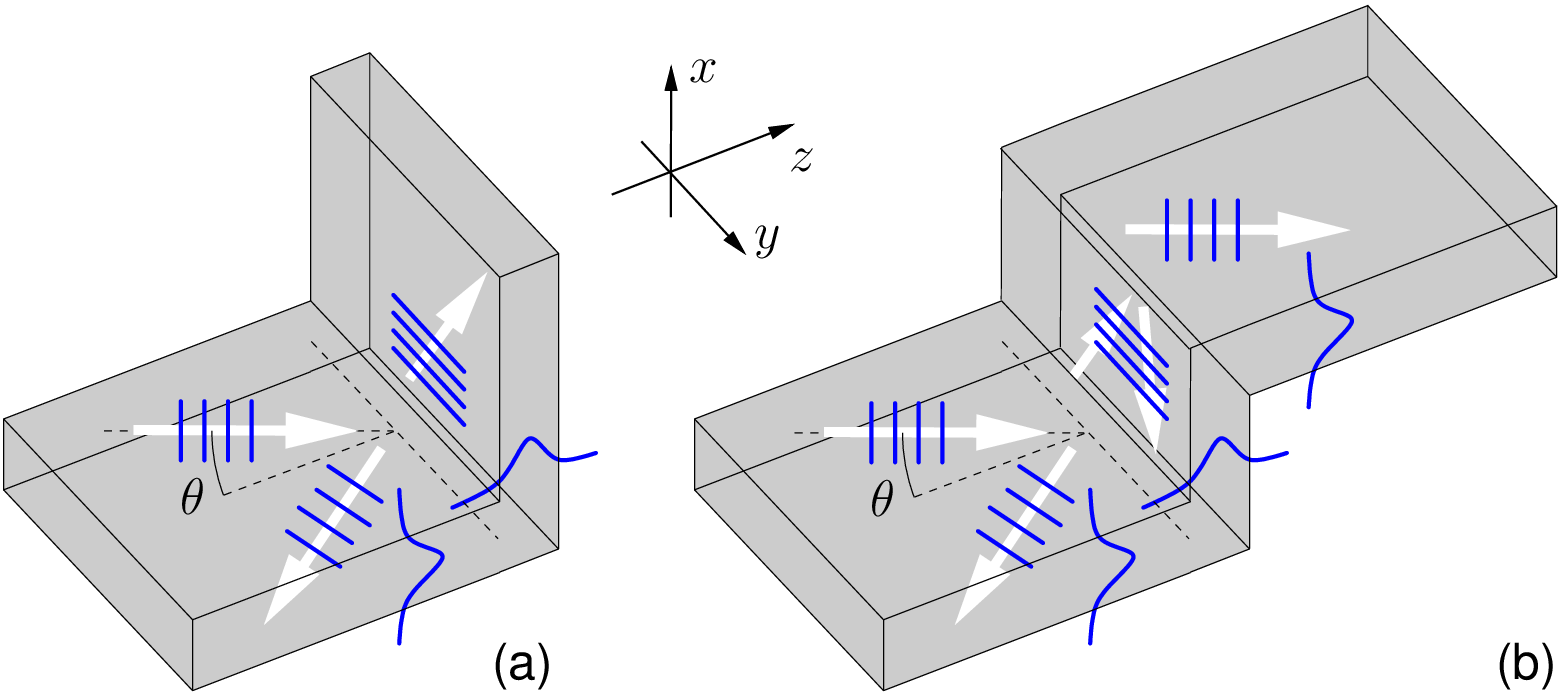}{10cm}{Oblique incidence of vertically guided, laterally unguided plane waves on 
a $90^\circ$ corner in a slab (a), and on a step-discontinuity consisting of two of 
these corners (b); incidence at angle $\theta$. 
These are meant to be $\pm y$-$\pm z$-infinite slabs, folded once (a) or twice (b).
}

\refsec{disc} introduces a rather general setting of a slab waveguide ``discontinuity'',
comprising an interior region with in principle arbitrary permittivity, that connects 
half-infinite slab waveguides at arbitrary positions and angles, but aligned such that 
the entire structure is homogeneous along one lateral coordinate (the $y$- axis in \reffig{kdk3Dsk}).  
A general form of Snell's law  applies to pairs of slab modes supported by 
these access channels. Depending on the modal properties of the incoming and outgoing 
slabs, one identifies critical angles of incidence that border on regimes with vanishing 
radiative losses, or on regimes with single-polarization or single mode propagation. 
Constituting the basis for early concepts of integrated optics \cite{ULM71,Tie77},  
this effect has been known for more than four decades (cf.\ e.g.\ Refs.\ \cite{SWM87,BiL90,MAF94,CTT03,Civ14}), 
but, to the best of our knowledge, has never been applied to configurations other 
than simple waveguide facets or transitions between co-aligned slabs with 
different cross sections, in particular not to configurations with orthogonal access 
slabs, such as for the present corners or steps. 

In \reftwosecs{kink}{dkink} we specialize this to high contrast silicon\,/\,silicon-oxide slabs.
The numerical simulations rely on the vectorial implementation of   
a scheme for 2-D quadridirectional eigenmode propagation (vQUEP) as described in 
Ref.\ \cite{Ham15}, based on a former scalar QUEP-variant \cite{Ham04}, and building on the 
bidirectional concepts of Refs.\ \cite{SWM87,BiL90}.
To investigate in how far the concepts extend to a more practical 3-D setting with 
at least weak lateral confinement, the propagation of semi-guided, laterally wide Gaussian 
wave bundles is considered in \refsec{bundles}. The vQUEP solver \cite{Ham15} has been 
extended accordingly.
A preliminary account of the present findings, highlighting the mechanism for 
full-transmission across the step, has been given in Ref.\ \cite{HHF15s}.

For the present paper our primary interest is in the
fundamental effect itself. However, structures as discussed here could e.g.\ enable power transfer between 
layers at different vertical levels in an 3-D integrated optical environment, 
e.g.\ in a framework of silicon photonics \cite{Sor06,KIJ06}.
Beyond the standard evanescent wave interaction \cite{SCN91} between the layers,
leading to devices measured in centimeters, for
vertical distances below 250\nm\ \cite{SCN91}, or to shorter devices, but for 
a vertical separation of merely a few tens of nanometers \cite{DKB05}, other concepts 
for vertical coupling include specifically tapered core shapes \cite{SBP08,BDH13},
radiative power transfer through grating couplers \cite{DOK04}, 
or even resonant interaction through vertically stacked microrings \cite{BeA13}.
We believe that the concepts from the present paper could offer here a 
viable, simple and robust alternative, in particular for situations 
where large lateral extensions of optical channels can be afforded, or 
if entire layers at different levels need to be uniformly ``flooded'' by light,
e.g.\ for purposes of optical pumping of active devices.


\section{Slab waveguide discontinuities at oblique incidence}
\labsec{disc} 

We start with an abstract look at a ``discontinuity'' in a slab waveguide, as introduced in 
\reffig{gcsk}. The structure comprises the slab that supports the incoming and possible reflected waves,
possible further slabs that support outgoing waves, and a central region with --- what concerns the arguments 
in these paragraphs --- arbitrary properties, hinted at by the dark patch in panel \reffign{gcsk}(a). 
In case of the configurations of \reffig{kdk3Dsk}, that region covers the actual 
corner, or the corners and the vertical segment, respectively. 
The entire structure is supposed to be 
constant along the lateral $y$-axis; we choose the $z$-direction as the axis of the incoming waveguide. 

\FIGh{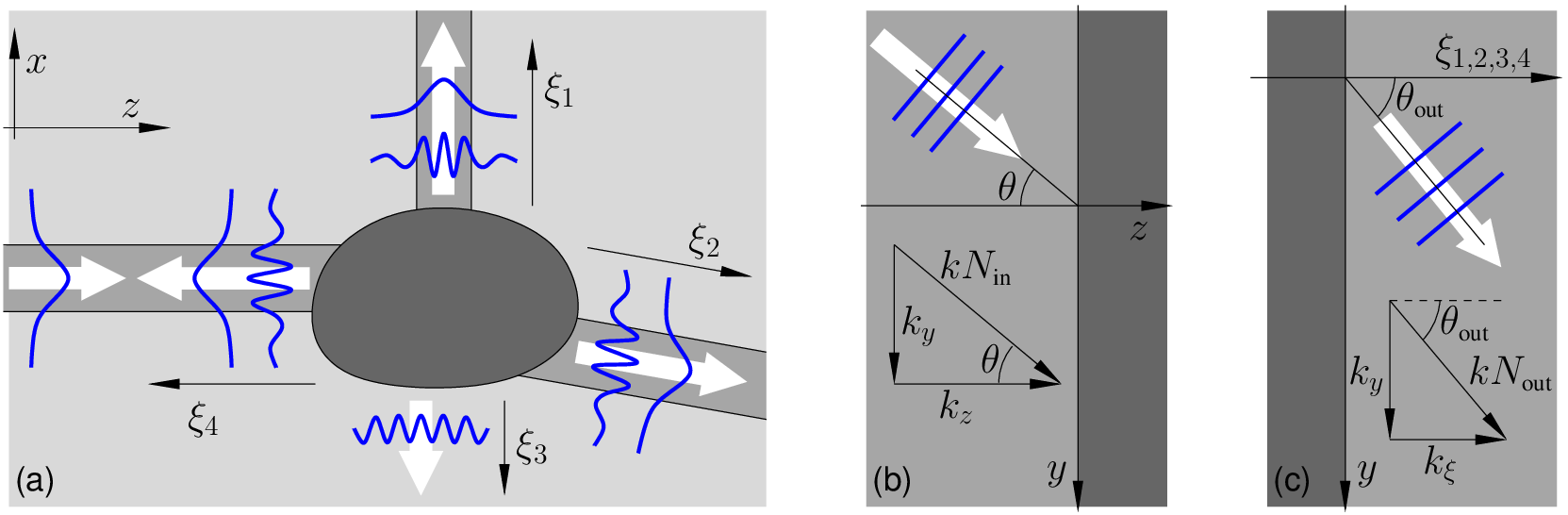}{0.9\textwidth}{Oblique incidence of semi-guided plane waves on 
a general ``discontinuity'', the darker central region; (a) cross-section (side-) view, (b) top view of the input waveguide, and 
(c) generic ``top view'' of any of the output slabs. Cartesian coordinates $x$, $y$, $z$ are 
oriented such that the incoming wave propagates with effective mode index $N\dn{in}$ at an angle $\theta$ in the horizontal 
$y$-$z$-plane, with a mode profile that depends on the vertical $x$-coordinate. 
Outgoing waves with effective mode index $N\dn{out}$ are observed at angles $\theta\dn{out}$ propagating in the $y$-$\xi$-planes, 
where $\xi$ represents a local coordinate along the slab core in the $x$-$z$-cross-sectional plane. 
The structure is constant along the $y$-axis. 
}

This concerns time harmonic fields $\sim \exp(\ie \omega t)$ with angular frequency $\omega = k\PCc = 2\pi\PCc/\lambda$, 
for vacuum wavenumber $k$, vacuum speed of light $\PCc$, and vacuum wavelength $\lambda$. For the incoming semi-guided 
wave we assume a polarized guided mode with vectorial profile $\nv{\Psi}\dn{in}(k_y; x)$ (see Refs.\ \cite{BiL90,Ham15}) and effective mode index $N\dn{in}$,  
supported by the incoming slab, propagating at an angle $\theta$ in the $y$-$z$-plane 
(cf.\ \reffig{gcsk}(b), also \refmfig{kdk3Dsk}, \reffign{ksk}, \reffign{dksk}). The incoming wave thus relates to a field dependence 
\beq{inc}
\sim \nv{\Psi}\dn{in}(k_y, x)\, \exp(-\ie (k_y y + k_z z)),
\eeq
with $k_y = k N\dn{in} \sin\theta$, $k_z = k N\dn{in} \cos\theta$, and $k_y^2 + k_z^2 = k^2N\dn{in}^2$. 
Due to the homogeneity of the problem along $y$, this harmonic $y$-dependence applies to all electromagnetic 
fields, at all positions; i.e.\ the global solution can be restricted to a single spatial Fourier component with
wavenumber $k_y$, here given by the angle of incidence $\theta$.

Next we look at a particular outgoing mode with profile $\nv{\Psi}\dn{out}$ and effective index $N\dn{out}$. 
As hinted at in \reffig{gcsk}(a), this can be an actual guided mode supported by one of the slabs with 
coordinates $\xi_1$, $\xi_2$, or $\xi_4 = -z$, but just as well some non-guided, radiated wave propagating in the 
$y$-$\xi_3$-plane. The outgoing field (cf.\ \reffig{gcsk}(c)) is of the form
\beq{out}
\sim \nv{\Psi}\dn{out}(k_y, \,.\,)\, \exp(-\ie (k_y y + k_{\xi} \xi)),
\eeq
where the wave equation \cite{Ham15} requires $k^2 N\dn{out}^2 = k_y^2 + k_{\xi}^2$ for the cross-sectional wavenumber  
$k_{\xi}$, still with $k_y = k N\dn{in} \sin\theta$ fixed by the incoming field. 

Depending on the angle of incidence, and specifically for each outgoing mode, one thus has to distinguish between
two cases. If the effective mode index $N\dn{out}$ is sufficiently large, i.e.\ if 
$k^2N\dn{out}^2 > k_y^2$, the outgoing field propagates at an angle $\theta\dn{out}$ with 
wavenumber $k_\xi = kN\dn{out} \cos\theta\dn{out}$, where the effective indices and the angles of incidence and refraction 
are related by Snell's law:
\beq{snellius}
N\dn{out}\sin\theta\dn{out} = N\dn{in}\sin\theta.
\eeq
Depending on the properties of the supporting slabs, outgoing waves are thus observed each at 
its own specific angle.

If, on the other hand, the effective mode index $N\dn{out}$ is smaller, i.e.\ if
$k^2N\dn{out}^2 < k_y^2$, the outgoing field becomes evanescent  
with an imaginary wavenumber $k_\xi = -\ie \sqrt{k_y^2 - k^2N\dn{out}^2}$. For the cross-sectional 
problem, these waves decay with growing distance $\xi$. In particular, these evanescent 
outgoing fields {\em do not carry optical power} \cite{LoS01}
(but note that they contribute significantly to the total field around the discontinuity). 

An increase of the angle of incidence $\theta$, starting from normal incidence $\theta=0$, 
can thus cause a change of a mode's type from $\xi$-propagating to $\xi$-evanescent. 
For an outgoing mode with effective index $N\dn{out} < N\dn{in}$, this happens if 
$k^2 N\dn{out}^2 = k^2 N\dn{in}^2\sin^2\theta$. Hence, for given input field and for each 
individual outgoing mode, by 
\beq{thetacr}
\sin\theta\dn{cr} = N\dn{out}/N\dn{in}
\eeq
one can define a characteristic angle $\theta\dn{cr}$, such that this outgoing mode   
does not carry power for incidence at $\theta > \theta\dn{cr}$ beyond that critical angle.

With a view to the corners and steps of \reffig{kdk3Dsk}, we now consider a structure where simple 
symmetric three-layer dielectric slabs, with core and cladding refractive indices $n\dn{g}$ and $n\dn{b}$,
constitute the (identical) access waveguides (cf.\ \refmfig{ksk}, \reffign{dksk}). The dimensions are assumed to be such 
that merely the fundamental modes of both polarizations, with effective indices $N\dn{TE0}$ and $N\dn{TM0}$,  
are guided, where  
$n\dn{g} > N\dn{TE0} > N\dn{TM0} > n\dn{b}$. The structure is excited by the fundamental TE mode.
In line with the former arguments, one can then conclude:
\begin{itemize}
\item
All modes of these waveguides that relate to radiative waves, with oscillatory behaviour in the 
cladding regions (``cladding modes''), have effective indices below the upper limit $n\dn{b}$ of the radiation continuum. 
Their characteristic angles \refeqn{thetacr} are smaller than the critical angle $\theta\dn{b}$, 
defined by $\sin\theta\dn{b} = n\dn{b}/N\dn{TE0}$, associated with the background refractive index. 
Consequently {\em all radiation losses vanish for incidence at angles $\theta > \theta\dn{b}$}. 
\item
All TM polarized modes supported by these waveguides have effective mode indices below $N\dn{TM0}$. 
Their characteristic angles \refeqn{thetacr} are thus smaller than the critical angle $\theta\dn{m}$,
defined by $\sin\theta\dn{m} = N\dn{TM0}/N\dn{TE0}$, associated with the fundamental TM wave. 
Consequently, {\em for incidence at angles $\theta > \theta\dn{m}$, all incoming optical power 
is carried away by outgoing fundamental guided TE waves}. 
\end{itemize}
Note that these arguments, as well as the former less specific statements, rely solely on 
the modal properties \refeqn{inc}, \refeqn{out} of the 
access waveguides, i.e.\ the reasoning applies to configurations with --- in principle ---
arbitrary interior and arbitrary extension of the region that connects the 
slab waveguide outlets. 

\subsection{Formal problem}
\labsubsec{formal}

We complement these more general considerations with a brief look at the rigorous equations, 
as discussed in Refs.\ \cite{CHH14,Ham15}. The problem is governed by the Maxwell curl equations in the 
frequency domain for the electric field $\sv{E}$ and magnetic field $\sv{H}$, for uncharged dielectric, 
nonmagnetic linear media with relative permittivity $\epsilon = n^2$, for vacuum permittivity $\epsilon_0$ 
and permeability $\mu_0$:  
\beq{curl}
\curl\sv{E} = - \ie \omega \mu_0\sv{H},\qquad \curl\sv{H} = \ie \omega\epsilon\epsilon_0\sv{E}.
\eeq
The properties of a $y$-homogeneous structure $\partial_y \epsilon = 0$, and of 
a corresponding harmonic $y$-dependence  
\beq{yharmon}
\displaystyle \left(\!\!\begin{array}{c} \sv{E} \\ \sv{H} \end{array}\!\!\right)
(x, y, z) =
\left(\!\!\begin{array}{c} \nv{E} \\ \nv{H} \end{array}\!\!\right)(x, z) \,\exp(-\ie k_y y)
\eeq 
of all fields, 
with the wavenumber $k_y = kN\dn{in} \sin\theta$ given by the incident wave, 
then lead to a system of vectorial equations on the $x$-$z$-cross section plane, 
here formulated for the two ``transverse'' electric field components,
\beq{formal}
\left(\!\!\begin{array}{cc}
\partial_x\displaystyle\frac{1}{\epsilon} \partial_x\epsilon + \partial_z^2 & \partial_x\displaystyle\frac{1}{\epsilon}\partial_z\epsilon -\partial_z\partial_x \\
\partial_z\displaystyle\frac{1}{\epsilon}\partial_x\epsilon - \partial_x\partial_z & \partial_x^2 + \partial_z\displaystyle\frac{1}{\epsilon}\partial_z\epsilon
\end{array}\!\!\right)
\left(\!\!\begin{array}{c} E_x \\ E_z \end{array}\!\!\right)
+
k^2\epsilon\dn{eff}
\left(\!\!\begin{array}{c} E_x \\ E_z \end{array}\!\!\right)
= 0,
\eeq
with an effective permittivity $\epsilon\dn{eff}$ depending on the angle of incidence:
\beq{effperm}
\epsilon\dn{eff}(x, z) = \epsilon(x, z) -N\dn{in}^2\sin^2\theta.
\eeq
\refmeq{formal}, \refeqn{effperm} are to be solved on a 2-D domain, with boundary conditions 
that are transparent for outgoing guided and nonguided waves, and that can accommodate the given
influx. Note that the problem coincides formally with the equations for the modes of channel 
waveguides with 2-D cross sections. The wavenumber $k_y$ appears in place of the channel mode propagation constant. 
Here, however, the equations need to be solved for the nonstandard  
boundary conditions, and need to be treated as a nonhomogeneous problem with the influx as a right-hand side,
not as an eigenvalue problem as in the case of channel mode analysis.

Details of the quasi-analytical solver (vQUEP, \cite{Metric}), that has been applied to 
generate the numerical results of this paper, can be found in Refs.\ \cite{Ham04,Ham15}.
In all cases the numerical parameters have been selected such that results are converged 
on the scale of the figures as given, where the power balance (conservation of energy) serves as one of the
criteria for convergence. With the exception of the 
configurations with pronounced radiative losses (small angles of incidence in \reffig{kascan}), relatively tight 
computational windows are sufficient, still with transparent-influx-boundary-conditions that incorporate the 
incoming waves (other non-guided, radiated fields are suppressed). Similar to guided mode analysis, also here the optical fields are well confined 
around the guiding cores. 

The vanishing of radiative losses for wave incidence beyond the critical angle can be understood
just as well in terms of these formal equations \cite{CHH14}. In regions with local constant permittivity
$\partial_x \epsilon = \partial_z \epsilon =0$, \refmeq{formal} reduce to the scalar Helmholtz 
equation
$\left( \partial_x^2 + \partial_z^2\right) \phi + k^2\epsilon\dn{eff} \phi = 0$, valid separately 
for all components $\phi = E_j, H_j$ of the optical electromagnetic fields, with the  
angle-dependent effective permittivity \refeqn{effperm}. Wave incidence beyond the critical angle 
$\theta\dn{b}$ then leads to a negative local effective permittivity, and consequently to evanescent 
wave propagation, in the cladding regions with refractive index $n\dn{b}$. 

The problem \refeqn{formal} should be distinguished from the standard 2-D Helmholtz problems 
for TE and TM waves, where the global solutions can be represented by a principal scalar field.
The latter emerge from \refmeq{curl} if both the structure and the solutions are assumed to be constant along 
one coordinate axis. In our case that corresponds to normal wave incidence $\theta = 0$.
Still the familiar scalar TE- and TM-slab modes constitute local solutions of \refeq{formal}, 
if their vectorial mode profiles are properly rotated \cite{BiL90,Ham15},
and if their cross sectional propagation constants $k_z$, $k_{\xi}$ are modified according 
to the global wavenumber $k_y$.


\section{Corner discontinuity}
\labsec{kink} 

Next we specialize to the corner configurations. \reffig{ksk} introduces  
a set of parameters that relate to silicon thin film cores in 
a silicon-oxide background at a typical near infrared wavelength.

\FIGv{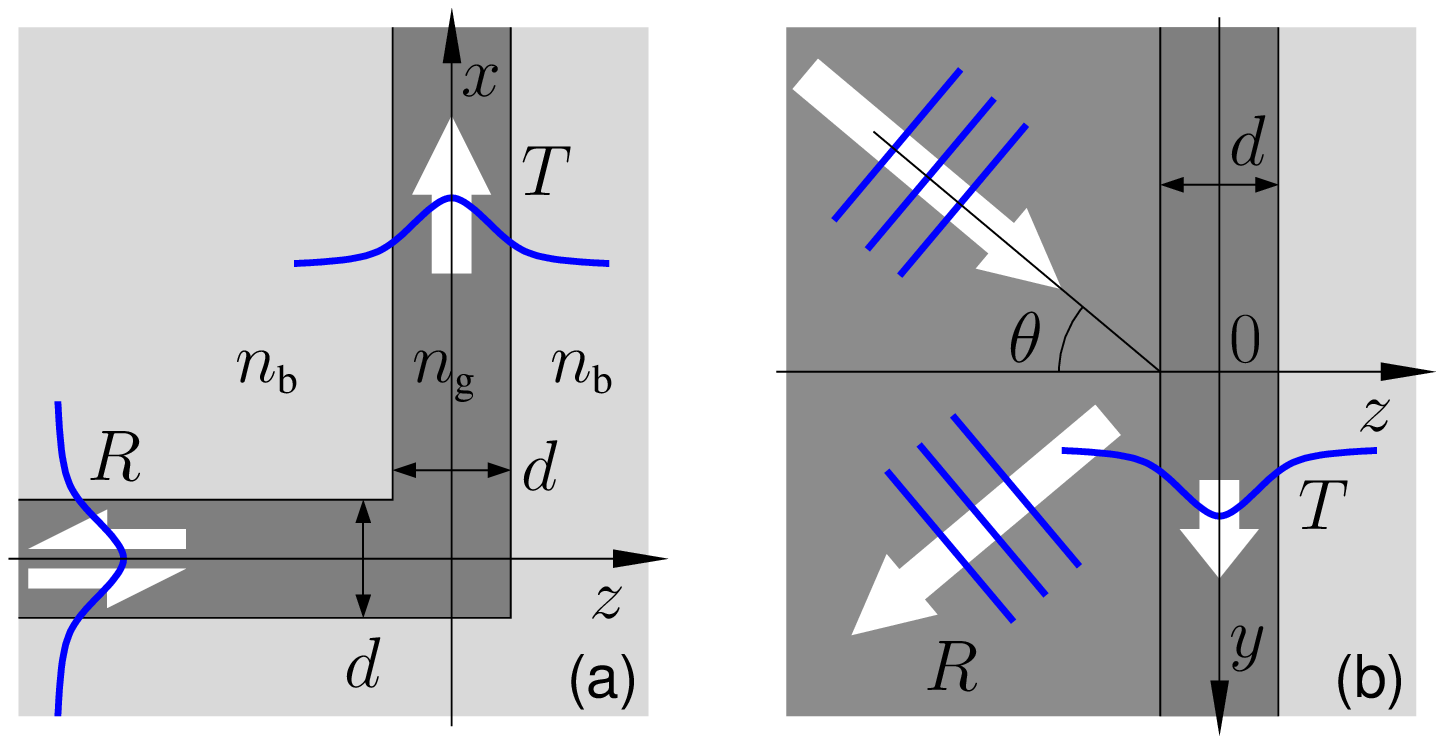}{7cm}{Slab waveguide with a straight fold, 
incidence of vertically ($x$) guided, laterally ($y$, $z$) unguided plane waves at angle $\theta$, 
cross-section (a) and top view (b). 
Parameters: incoming TE-polarized light at vacuum wavelength $\lambda = 1.55\mum$,
refractive indices $n\dn{g} = 3.4$ (guiding regions), $n\dn{b} = 1.45$ (background), core thickness $d = 0.25\mum$.
}

\refmfig{kascan}, \reffign{kflds}, and \reftab{rtvals} collect the results of a series of 
rigorous vQUEP simulations for the corner structures. \reffig{kascan} shows the variation of the  
reflectances $R\dn{TE}$, $R\dn{TM}$ and transmittances $T\dn{TE}$, $T\dn{TM}$ 
of the fundamental polarized modes with the angle of incidence. 
As to be expected, only moderate levels of reflection and transmission
(cf.\ \reftab{rtvals}) are observed at normal incidence $\theta=0$,
with most of the optical power being lost to radiated fields.
Pronounced radiative losses are also present at larger angles below $\theta\dn{b}$, 
where similar amounts of power are carried upwards by TE- and TM-polarized waves.  

\FIGv{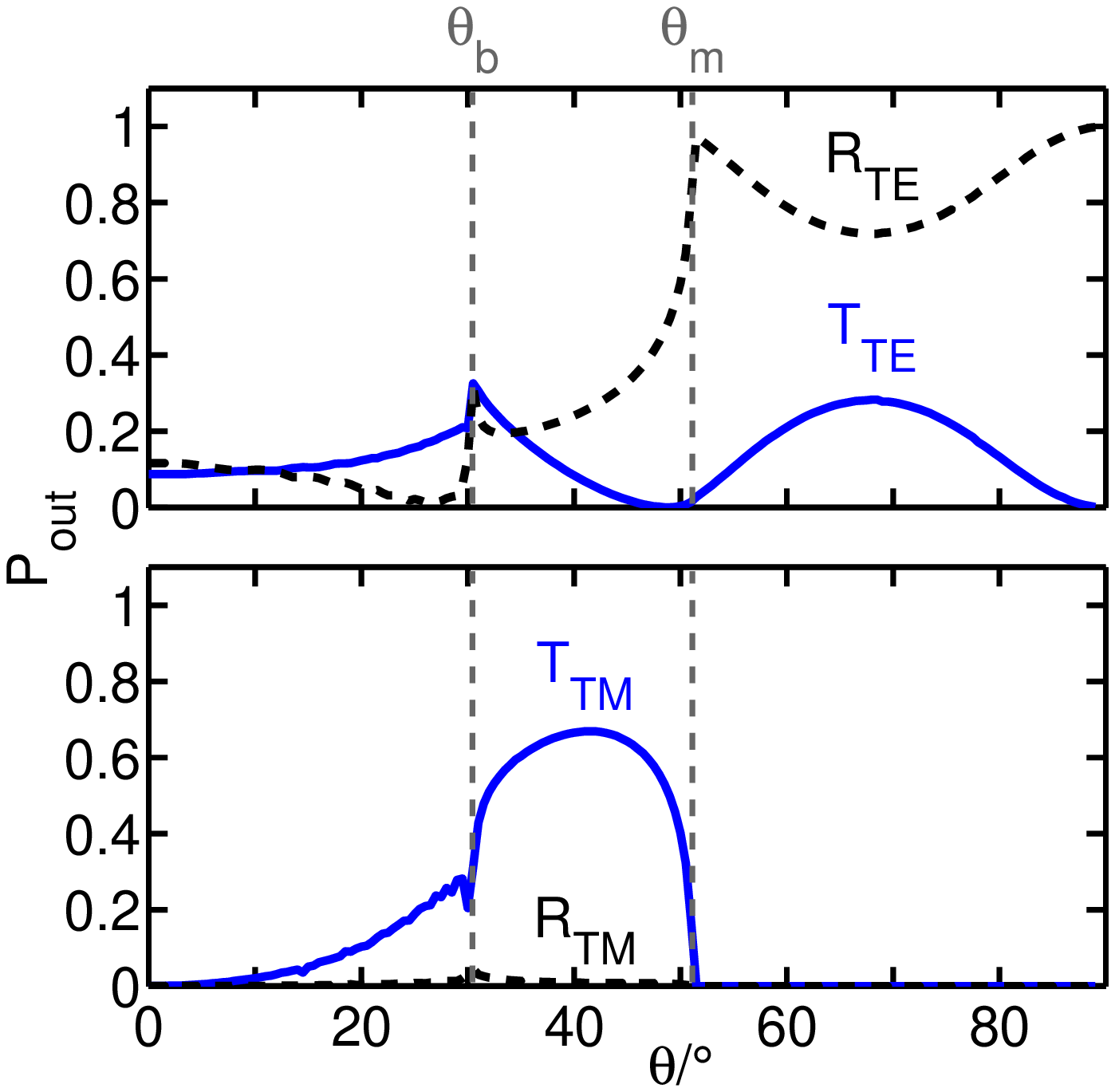}{6.5cm}{For the corner of \reffig{ksk}: relative reflected ($R$, dashed) and upwards transmitted ($T$, solid) guided power carried by 
TE (upper panel) and TM polarized waves (lower panel), as a function of the angle of incidence $\theta$. 
The critical angles $\theta\dn{b} = 30.45^\circ$ 
and $\theta\dn{m}=51.14^\circ$ are indicated.
}

As predicted in \refsec{disc}, radiation losses vanish entirely $R\dn{TE}+R\dn{TM}+T\dn{TE}+T\dn{TM} = 1$ 
for $\theta > \theta\dn{b}$, beyond the critical angle $\theta\dn{b}$ for propagating waves in the cladding.
Power conversion to outgoing TM waves reaches a strong maximum in between $\theta\dn{b}$ and $\theta\dn{m}$, and is 
suppressed entirely beyond the critical angle $\theta\dn{m}$ for propagating TM waves, where all outgoing 
waves are purely TE polarized: $R\dn{TE} +T\dn{TE} = 1$ for $\theta > \theta\dn{m}$.
For even higher angles of incidence one observes a maximum in the TE transmission, which vanishes 
for grazing incidence as $\theta$ approaches $90^\circ$. 

\FIGh{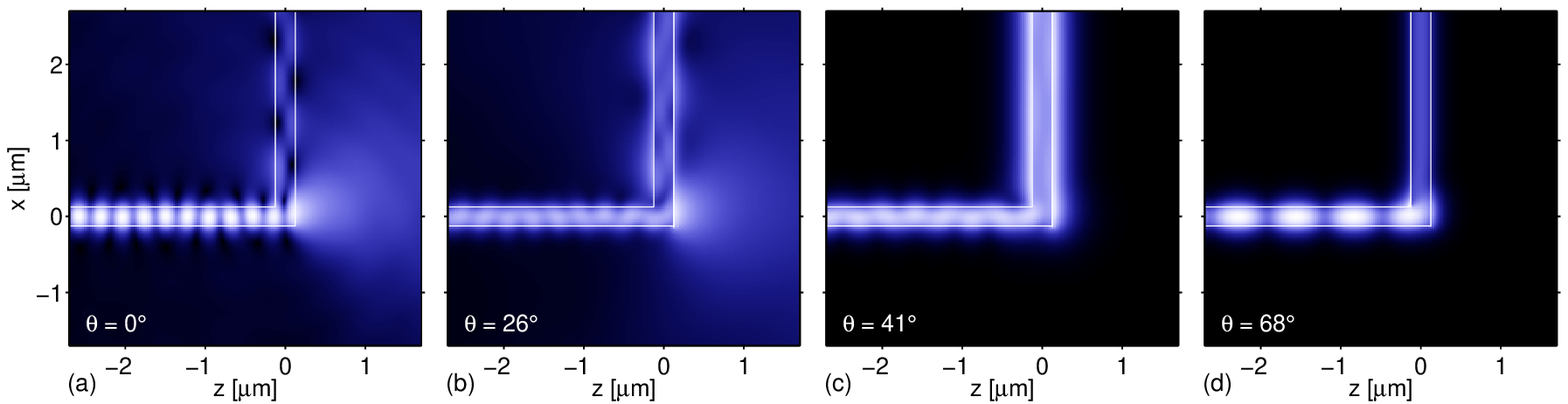}{\textwidth}{Field profiles $|\nv{E}|$ for the corner discontinuity of \reffig{ksk}, 
for angles of incidence $\theta = 0^\circ, 26^\circ, 41^\circ, 68^\circ$.  
}

\reffig{kflds} provides a few examples of fields observed for these typical cases. The 
plots show absolute values of the electric field vector $|\nv{E}|$, with the color scales 
adjusted such that variations in regions with small field strengths become visible.
Note that, irrespective of the critical angles, these rigorous solutions contain contributions from 
the complete sets of local vectorial slab modes, of both polarizations, which are used as 
the expansions bases \cite{Ham15}. If beyond a related critical angle, however, the then evanescent 
modes stay localized around the corner region, not contributing to the ``far-fields'' in the outlets.

While pronounced radiative losses are evident in parts (a) and (b) for $\theta < \theta\dn{b}$, all fields 
remain confined around the core regions in (c) and (d) for angles $\theta > \theta\dn{b}$. 
Outgoing TM waves lead to strong electric fields immediately 
outside the cores of the vertical channel (``highlighted'' core edges) in panels (b) and (c), but are 
absent for the scalar problem at normal incidence (a) and for $\theta > \theta\dn{m}$ in (d).  

\reftab{rtvals} compares values for reflectances and transmittances, together with selected 
parameters, for the configurations of all field plots in this paper.
Although here we consider the corners merely as building blocks for the steps  
of the next sections, a lossless corner structure as the one of \reffig{kflds}(c), that  
channels a total of $74\%$ of the horizontal input into the vertical, might be of practical 
interest in its own. 

\TABh{rtvals}{%
\setlength{\tabcolsep}{1mm}%
\begin{tabular*}{\textwidth}[b]{@{}@{\extracolsep{\fill}}l|llllll|llllll|c@{}}
~                    & \multicolumn{6}{c|}{corner}                                                 & \multicolumn{6}{c|}{step}                                                   & u-turn, bridge, stair, s-bend  \\ \hline
$\theta$, $\theta_0$ & $ 0^\circ$ & $26^\circ$ & $41^\circ$ & $41^\circ$ & $68^\circ$ & $68^\circ$ & $41^\circ$ & $41^\circ$ & $41^\circ$ & $68^\circ$ & $68^\circ$ & $68^\circ$ & $68^\circ$                     \\
$h/\mum$             & n/a        & n/a        & n/a        & n/a        & n/a        & n/a        & $1.83$     & $1.83$     & $1.83$     & $2.15$     & $2.15$     & $2.15$     & $2.15$                         \\
$W_y/\mum$           & $\infty$   & $\infty$   & $\infty$   & $13$       & $\infty$   & $27$       & $\infty$   & $13$       & $60$       & $\infty$   & $27$       & $481$      & $\infty$                       \\ 
$W\dn{cr}/\mum$      & $\infty$   & $\infty$   & $\infty$   & $10$       & $\infty$   & $10$       & $\infty$   & $10$       & $45$       & $\infty$   & $10$       & $180$      & $\infty$                       \\ \hline
$R\dn{TE}$           & $0.12$     & $0.01$     & $0.25$     & $0.25$     & $0.72$     & $0.72$     & $<0.01$    & $0.20$     & $0.02$     & $<0.01$    & $0.66$     & $0.03$     & $<0.01$                        \\
$R\dn{TM}$           & $0$        & $0.01$     & $0.01$     & $0.01$     & $0$        & $0$        & $<0.01$    & $<0.01$    & $<0.01$    & $0$        & $0$        & $0$        & $0$                            \\
$T\dn{TE}$           & $0.09$     & $0.17$     & $0.07$     & $0.07$     & $0.28$     & $0.28$     & $0.97$     & $0.78$     & $0.96$     & $>0.99$    & $0.34$     & $0.97$     & $>0.99$                        \\
$T\dn{TM}$           & $0$        & $0.21$     & $0.67$     & $0.67$     & $0$        & $0$        & $0.02$     & $0.02$     & $0.02$     & $0$        & $0$        & $0$        & $0$                            \\ \hline
Figure              
& \reffign{kflds}(a)
& \reffign{kflds}(b)
& \reffign{kflds}(c)
& \reffign{kbflds}(a)
& \reffign{kflds}(d)
& \reffign{kbflds}(b)
& \reffign{dkflds}(a)
& \reffign{dkbflds}(a)
& \reffign{dkbflds}(b)
& \reffign{dkflds}(b)
& \reffign{dkbflds}(c)
& \reffign{dkbflds}(d)
& \reffign{sbtdflds}(a, b, c, d)
\end{tabular*}\par
}{%
Reflectance and transmittance values 
$R\dn{TE}$, 
$R\dn{TM}$, 
$T\dn{TE}$, 
$T\dn{TM}$ 
for the polarized fundamental slab modes / for wave bundles made of these, and varying parameters 
(angle of incidence $\theta$, vertical layer distance $h$, bundle widths $W_y$, $W\dn{cr}$ along $y$ and along the beam cross section)
for the configurations relating to the field plots in this paper.
}


\section{Step discontinuity}
\labsec{dkink} 

Combination of two of the former corners gives a step structure that connects  
optical layers at different elevations. \reffig{dksk} introduces coordinates and 
parameters. Aiming at efficient power transfer, we focus on two of the lossless
cases from \refsec{kink}, 
for $\theta=41^\circ$ (\reffig{kflds}(c), this will be named configuration C0)
and 
for $\theta=68^\circ$ (\reffig{kflds}(d), labeled configuration C1). 
Both relate to transmission maxima in \reffig{kascan}. For each case 
we assume that the two combined corners are identical. 

\FIGv{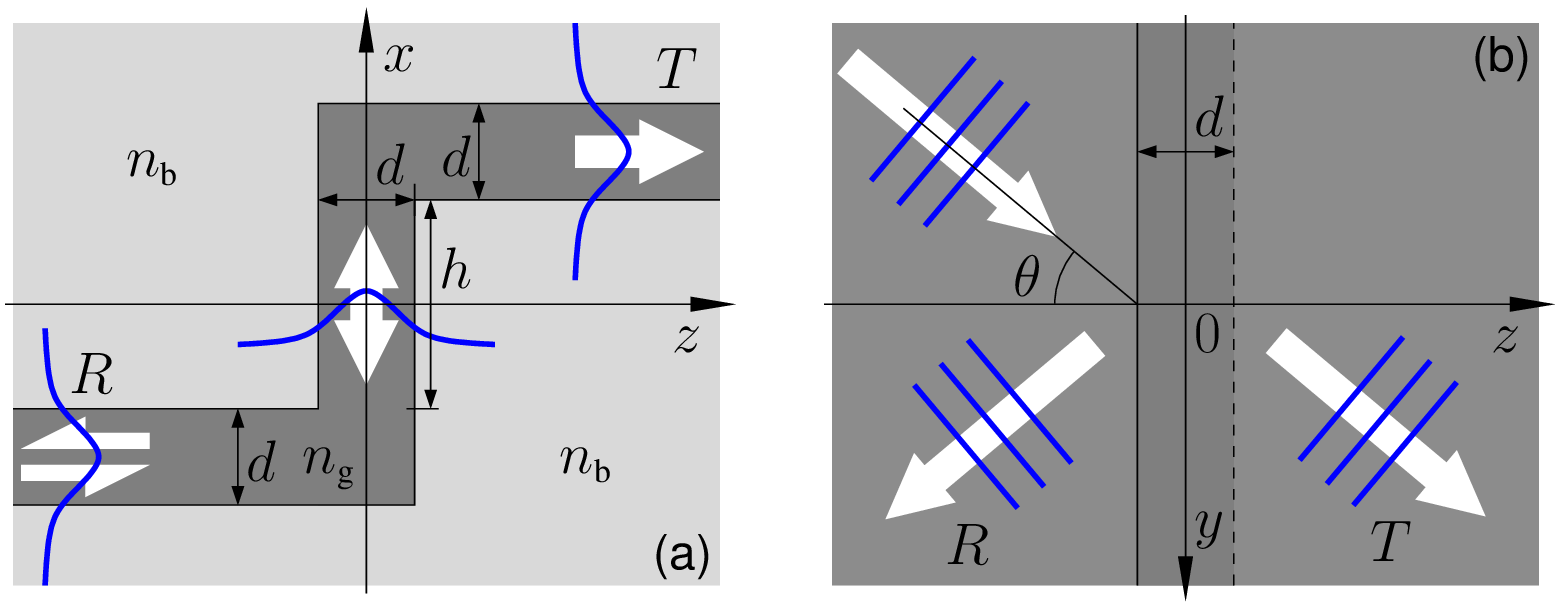}{9cm}{A step configuration, 
cross-section (a) and top view (b).  
Parameters are as given for \reffig{ksk}, with an additional value $h$ for the 
vertical core distance.
}

With the vertical distance $h$ between the horizontal layers, only one new parameter 
appears. Parts (a) and (c) of \reffig{dkscan} show the transmittance and reflectance of the step
as a function of this distance, calculated by rigorous vQUEP simulations for the 
full step structure. For both configurations, the angles of incidence exceed the 
critical angle $\theta\dn{b}$, hence the steps are lossless.

\FIGh{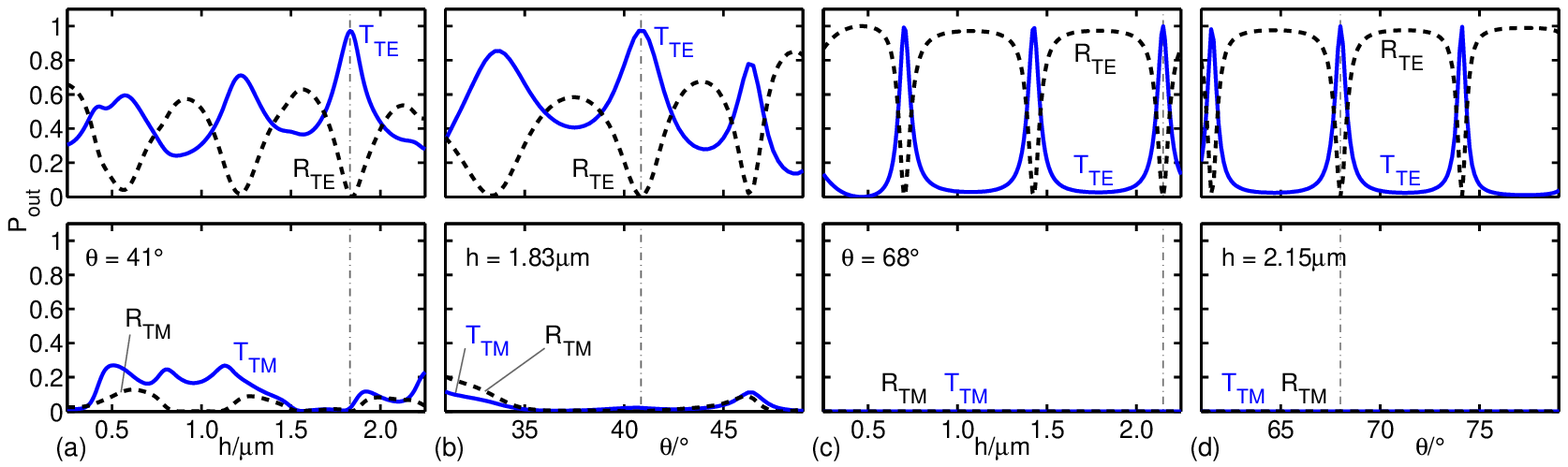}{\textwidth}{%
For the steps of \reffig{dksk}, TE-input: polarized modal transmittance 
($T\dn{TE}$, $T\dn{TM}$)
and reflectance 
($R\dn{TE}$, $R\dn{TM}$)
versus the step height $h$ (a, c) for given incidence angle $\theta$, and versus 
the angle of incidence for given step height (b, d). Thin dash-dotted lines indicate the 
values $\theta = 41^\circ, 68^\circ$ and $h = 1.83\mum, 2.15\mum$  
selected for the configurations (C0) and (C1) of \reffig{dkflds}.
}

We look at configuration (C1) first. Since the related corner structures (cf.\ \reffig{kflds}(d), entry in  
\reftab{rtvals}) transmit and reflect only TE waves, only the upward and downward propagating versions of 
the fundamental TE mode mediate between the two corners in the vertical part of the step. One might thus 
compare the step with a standard Fabry-Perot interferometer \cite{BoW99}, where the corners play the 
role of the (identical) partial reflectors. Consequently, the scan over the vertical separation $h$ reveals  
multiple-beam fringes resulting from the interference of the repeatedly up- and downwards reflected 
TE modes. Maxima with virtually full transmission appear for certain equidistant heights. 

\reffig{dkflds}(b) shows the field pattern for an optimum configuration with full TE-to-TE transmission 
over a vertical distance of $h=2.15\mum$. The field is confined around the core regions, without any 
beating pattern, i.e.\ with purely forward waves, in the horizontal segments, and with strong resonant, 
mostly standing waves in the vertical slab. 

\FIGh{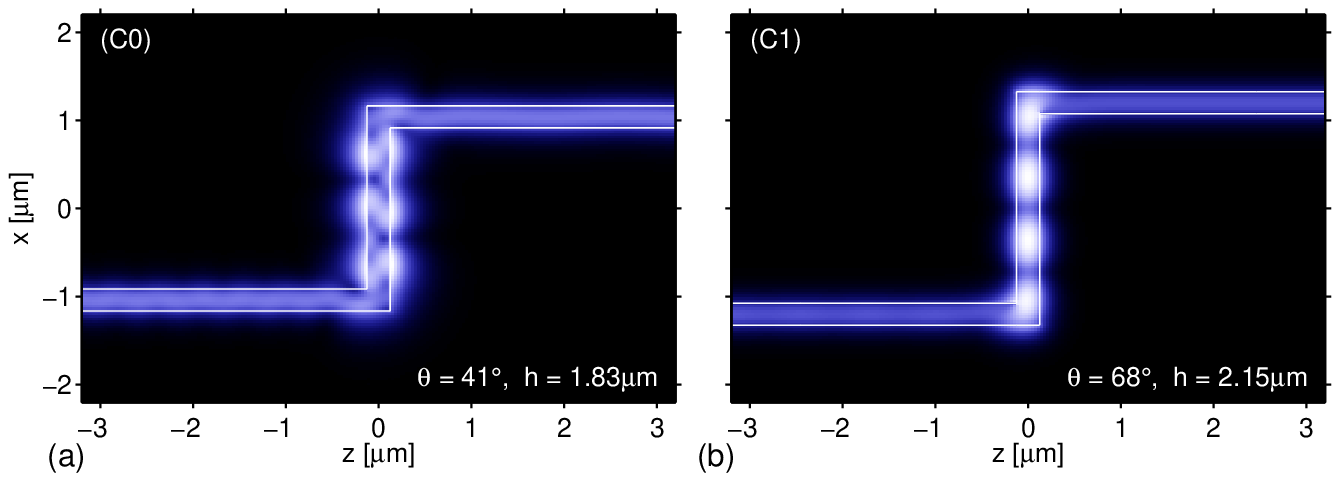}{0.8\textwidth}{Field profiles $|\nv{E}|$ for two step configurations 
as in \reffig{dksk}, with (almost) full transmission. For the fundamental guided slab modes 
one observes transmittance and reflectance levels as given in \reftab{rtvals}. 
}

Regarding configuration (C0), at $\theta=41^\circ$ and for TE input, the corners transmit the 
fundamental modes of both polarizations (with a dominant TM part, cf.\ the entry of \reffig{kflds}(c) in \reftab{rtvals}).
Hence, in a step structure made of these corners, one must expect that TM as well as TE waves play a role in the interference in the 
vertical segment. Accordingly, the scan over the vertical separation in \reffig{dkflds}(a) shows a much 
less regular dependence. Still, for a distance $h=1.83\mum$, a lossless configuration with virtually 
full transmission can be identified. The transmission is mostly TE polarized (reciprocity arguments 
apply \cite{Ham15}), with a TM contribution of about $2\%$. The field plot in \reffig{dkflds}(a)
shows a major TM contribution to the resonance in the vertical slab, visible through the strong 
electric field contributions next to the vertical core edges, similar to \reffig{kflds}(c).

In principle, given the properties of the constituting corners in the form of
scattering matrices, optical transmission through the step structures can be analyzed
analytically. For configuration (C1), where propagating TM waves are suppressed, 
the corners are represented by a $2 \times 2$ matrix that relates to incoming- and
outgoing TE waves. Configuration (C0) requires a $4 \times 4$ matrix to accommodate
TE as well as TM waves. The model would be applicable to steps of sufficient height,
where evanescent fields around the corner regions do not play a role.

As a resonant effect, the property of full transmission must be expected to be 
sensitive, to some degree, to all parameters that enter. With a view to \refsec{bundles} we select the input
angle for a further parameter scan. According to panels (b) and (d) of \reffig{dkscan}, 
although less perfect in peak performance, configuration (C0) might turn out to be 
more robust than (C1), here concerning variations of the angle of incidence.


\section{Semi-guided beams}
\labsec{bundles} 

All examples discussed in \refmsec{kink} and \refsecn{dkink} concern semi-guided 
plane waves that extend infinitely in the $\pm y$ direction. We now look at bundles of the former 
solutions, for a typically small range of angles of incidence, or of wavenumbers $k_y$, respectively,
with the aim of modeling the incidence of semi-guided, laterally wide but localized beams on 
the corners and steps. These are wave packets of the form
\beqnn
\left(\!\!\!\begin{array}{c} \nv{E} \\ \nv{H} \end{array}\!\!\!\right)\!(x,y,z) = 
A \int \exp\!\left(-\frac{(k_y - k_{y0})^2}{w_k^2}\right)
\eeqnn
\beq{bundle}
\mbox{~}\hspace{2.6cm}\cdot\,
\Big\{
\nv{\Psi}\dn{in}(k_y; x) \exp\left(-\ie k_z(k_y)(z-z_0)\right) +
\nv{\rho}(k_y; x, z)
\Big\}\,
\exp\left(-\ie k_y (y-y_0)\right)
\,\,\mbox{d}k_y,
\eeq
with a Gaussian weighting of half width $w_k$, centered around a primary wavenumber $k_{y0} = k N\dn{in} \sin \theta_0$,   
for primary angle of incidence $\theta_0$, and an (arbitrary) amplitude $A$. Phase factors have been introduced to position the focus in the 
$y$-$z$-plane at $(y_0, z_0)$. The central term in curly brackets represents the numerical 2-D vQUEP solution, formally 
separated into a contribution of the incoming field with mode profile $\nv{\Psi}\dn{in}$, and a remainder $\nv{\rho}$.
The integrals are evaluated by numerical quadrature \cite{PTV92}. We thus arrive at approximations of $y$-localized 
true 3-D solutions, as a basis for the results in \refmfig{kbflds} and \reffign{dkbflds}.

\FIGh{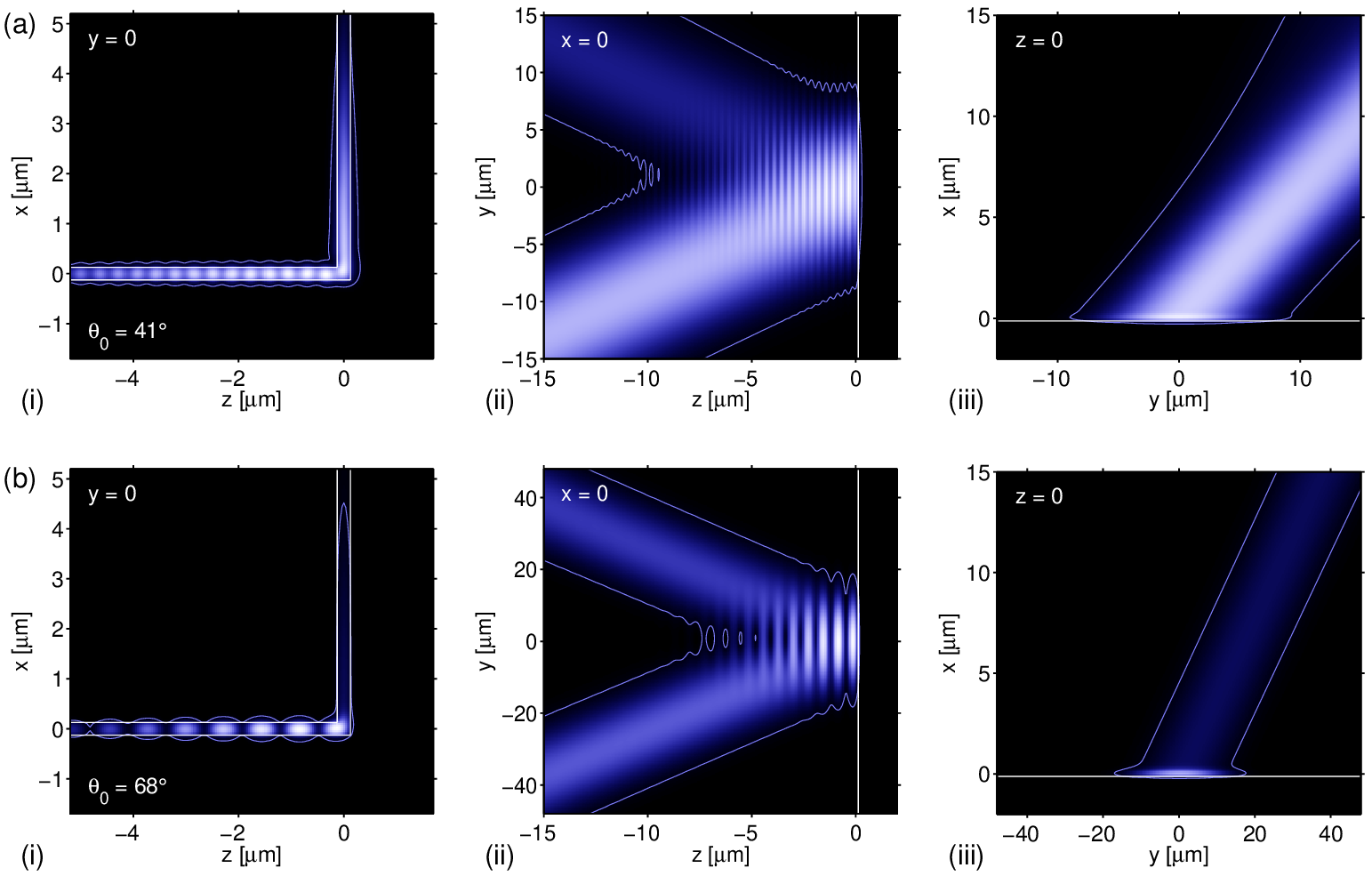}{\textwidth}{%
Optical electromagnetic energy density for incidence of Gaussian wave bundles (\refeq{bundle}) with a cross section width $W\dn{cr} = 10\mum$ on a corner discontinuity as in \reffig{ksk}, for 
principal angles of incidence (a) $\theta_0 = 41^\circ$ and (b) $\theta_0 = 68^\circ$. The panels show cross section views (i), bottom views (ii), and 
views of the plane through the center of the vertical slab (iii).  Further parameters, and transmittance and reflectance levels for the bundles are 
given in \reftab{rtvals}.
}

\FIGp{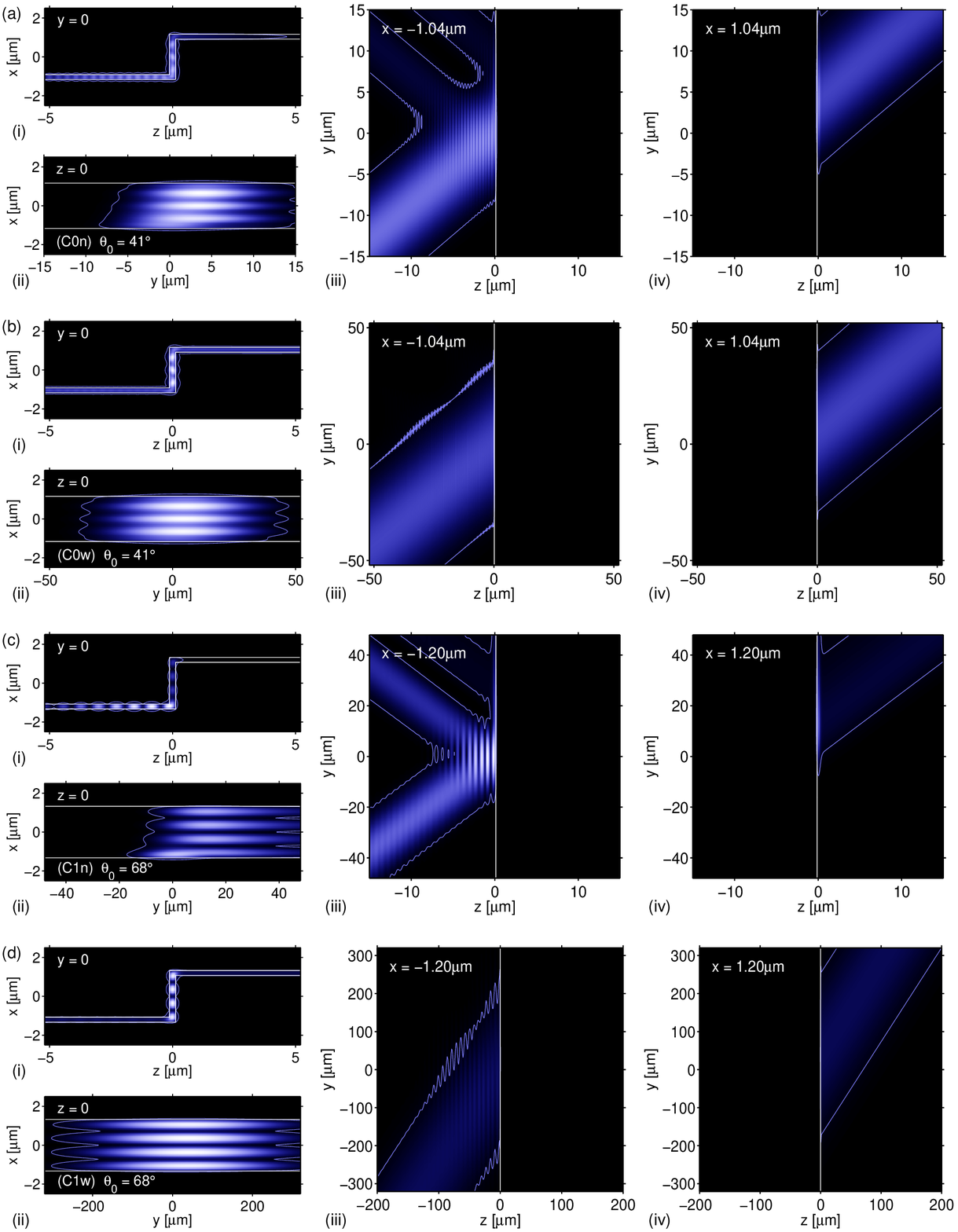}{\textwidth}{
Optical electromagnetic energy density for incidence of Gaussian wave bundles \refeqn{bundle} with cross section widths of   
$W\dn{cr} = 10\mum$ (C0n, a),
$W\dn{cr} = 45\mum$ (C0w, b),
$W\dn{cr} = 10\mum$ (C1n, c),
$W\dn{cr} = 180\mum$ (C1w, d),
on step discontinuities as in \reffig{dksk}, for 
principal angles of incidence $\theta_0 = 41^\circ$ (C0, a, b) and $\theta_0 = 68^\circ$ (C1, c, d). The panels show cross section views (i), 
views of the plane through the center of the vertical segment (ii), 
and bottom views at the levels of the lower (iii) and of the upper slab (iv). Modal transmittances and reflectances for the 
bundles are given in \reftab{rtvals}.
}

For a proper choice of wave packet parameters it is instrumental to evaluate the incident part 
(everything except $\nv{\rho}$) of \refeq{bundle} a little further. Assuming a small spectral width $w_k$, and neglecting the 
effect of the rotation of the vectorial mode profile (by replacing 
$\nv{\Psi}\dn{in}(k_{y}; x)$ by 
$\nv{\Psi}\dn{in}(k_{y0}; x)$),  
the incident field can be expressed as 
\clearpage
\beqnn
\left(\!\!\!\begin{array}{c} \nv{E} \\ \nv{H} \end{array}\!\!\!\right)\dn{in}\!(x,y,z) \approx 
\sqrt{\pi}\frac{2\,A}{W_y} 
\,\exp\!\!\left(-\frac{\left(\!\!(y - y_0)- \frac{k_{y0}}{k_{z0}}(z-z_0)\!\!\right)^2}{(W_y/2)^2}\right)
\eeqnn
\beq{iwcartc}
\mbox{~}\hspace{2.6cm}\cdot\,
\nv{\Psi}\dn{in}(k_{y0}; x) 
\,\exp\left(-\ie (k_{y0} (y-y_0) + k_{z0}(z-z_0))\right),
\eeq
with $k_{z0} = k N\dn{in} \cos \theta_0$. In the $y$-$z$-plane this is a Gaussian beam with fields of full 
width $W_y$ along $y$ at $1/\mathrm{e}$-level, where $W_y = 4/w_k$.
By introducing the cross section position $c$ and longitudinal position $l$, relative to the focus,  
as new coordinates through  
$y = y_0 + l \sin\theta_0 + c \cos\theta_0$, 
$z = z_0 + l \cos\theta_0 - c \sin\theta_0$, 
one can write the incoming field in the more succinct form
\beq{iwbeamc}
\left(\!\!\!\begin{array}{c} \nv{E} \\ \nv{H} \end{array}\!\!\!\right)\dn{in}\!(x,c,l) \approx 
\sqrt{\pi}\frac{2\,A}{W_y} \,\,
\exp\left(-\frac{c^2}{(W\dn{cr}/2)^2}\right)\,\,
\nv{\Psi}\dn{in}(k_{y0}; x) \,\,
\exp\left(-\ie k N\dn{in} l\right).
\eeq
Here the cross-section-width $W\dn{cr}$ (full width of the field at $1/\mathrm{e}$-level, at focus, in the direction perpendicular to 
the beam axis) is related to the $y$-width by $W\dn{cr} = W_y \cos\theta_0$. Both quantities might be practically relevant, hence 
both values are listed in \reftab{rtvals}.

\refmfig{kbflds} and \reffign{dkbflds} collect corresponding results for our corner and step structures.
The plots show the electromagnetic energy density $(\epsilon_0\epsilon|\nv{E}|^2+\mu_0|\nv{H}|^2)/4$ 
as a physically more relevant quantity. Uniform color scales have been adopted for all three or four panels that 
relate to different views of the same configuration. In some cases the strong resonant fields visible in the 
longitudinal views lead to rather dark incoming/outgoing beams (although these carry power of unit order). 
The single contour at $2\%$ of the energy density maximum is meant to better accentuate the form of these beams.

According to the values in \reftab{rtvals}, both corner configurations work 
almost identically for bundles with cross section width $W\dn{cr}=10\mum$  
and for incoming plane waves. The reflected and transmitted 
beams in \reffig{kbflds} remain nicely confined. This can be explained by the mere 
weak angular dependence of the transmission curves in \reffig{kascan} around the maxima 
selected for the corner configurations (C0) and (C1). 

In contrast, for the same wave bundles of width $W\dn{cr} = 10\mum$, the performance of the 
step structures deteriorates, when compared to the case of plane wave incidence. Here the strong 
angular dependence of the resonant transmission maxima in \reffig{dkscan} becomes relevant. 
The reflected beams in parts (a, iii) and (c, iii) of \reffig{dkbflds} show pronounced 
sidelobes. Merely the central part of the incoming bundle appears to be transmitted.
The effect is more pronounced for configuration (C1) than for (C0) due to the 
narrower transmission peak in \reffig{dkscan}(d).

Hence we choose wider input beams for both configurations as our last examples.
The cross section widths of $W\dn{cr}=45\mum$ (C0) and $W\dn{cr}=180\mum$ selected for 
panels \reffign{dkbflds}(b, d) roughly correspond to an angular range around $\theta_0$ with TE-transmission
above $90\%$, i.e.\ $w_k$, and $W_y$, $W\dn{cr}$, respectively, are adjusted such that 
$T\dn{TE}(\theta) > 90\%$ for all $\theta \in [\theta_0-\Delta\theta, \theta_0+\Delta\theta]$, 
for $\Delta\theta = w_k /(k N\dn{in} \cos\theta_0)$. According to \reftab{rtvals}, for incoming
beams this wide, the step configurations come reasonably close to their ideal plane-wave performance. 


\section{Concluding remarks}

Semi-guided plane waves at sufficiently high angles of incidence 
propagate across arbitrary straight slab waveguide 
discontinuities without any radiation losses. This effect can be understood 
in terms of a variant of 
Snell's law, that relates effective mode indices and angles of 
incidence and refraction in pairs of access channels. For a quantitative  
analysis of these configurations, the frequency-domain
Maxwell equations reduce to a vectorial 2-D system, to be solved 
on a computational window with transparent-influx boundary conditions. 
Our rigorous quasi-analytical solver enables the convenient analysis of 
structures with general rectangular permittivity distributions.
Simulations of high-contrast Si\,/\,SiO$_2$ corners 
lead to the identification of two example configurations with local transmission 
maxima. A Fabry-Perot-like resonance effect permits to combine two identical 
corners into step configurations with full transmission of the incoming 
semi-guided plane wave, i.e.\ with (numerically) ideal behaviour. 
Bundles of the former solutions can serve as examples of what happens
to semi-guided, laterally localized beams, when incident on the 
corners or step structures. Our examples show that transmission 
properties very close to the laterally infinite case can be realized 
with beams of sufficient width. 

Once a working step configuration has been identified, extension to further, 
perhaps also intriguing examples is obvious.
Due to the waveguide symmetry, the transmission properties of the corners is not affected if one changes their direction. This enables 
u-turn-like structures as in \reffig{sbtdflds}(a). Resonant vertical segments in the form of steps or u-turns can be connected by pieces of horizontal slabs
of arbitrary length. One is thus led to the bridge-, staircase-, or s-bend-like configurations of \reffig{sbtdflds}(b, c, d). Our vQUEP simulations predict 
transmittance levels $T>99\%$ in all cases.

\FIGv{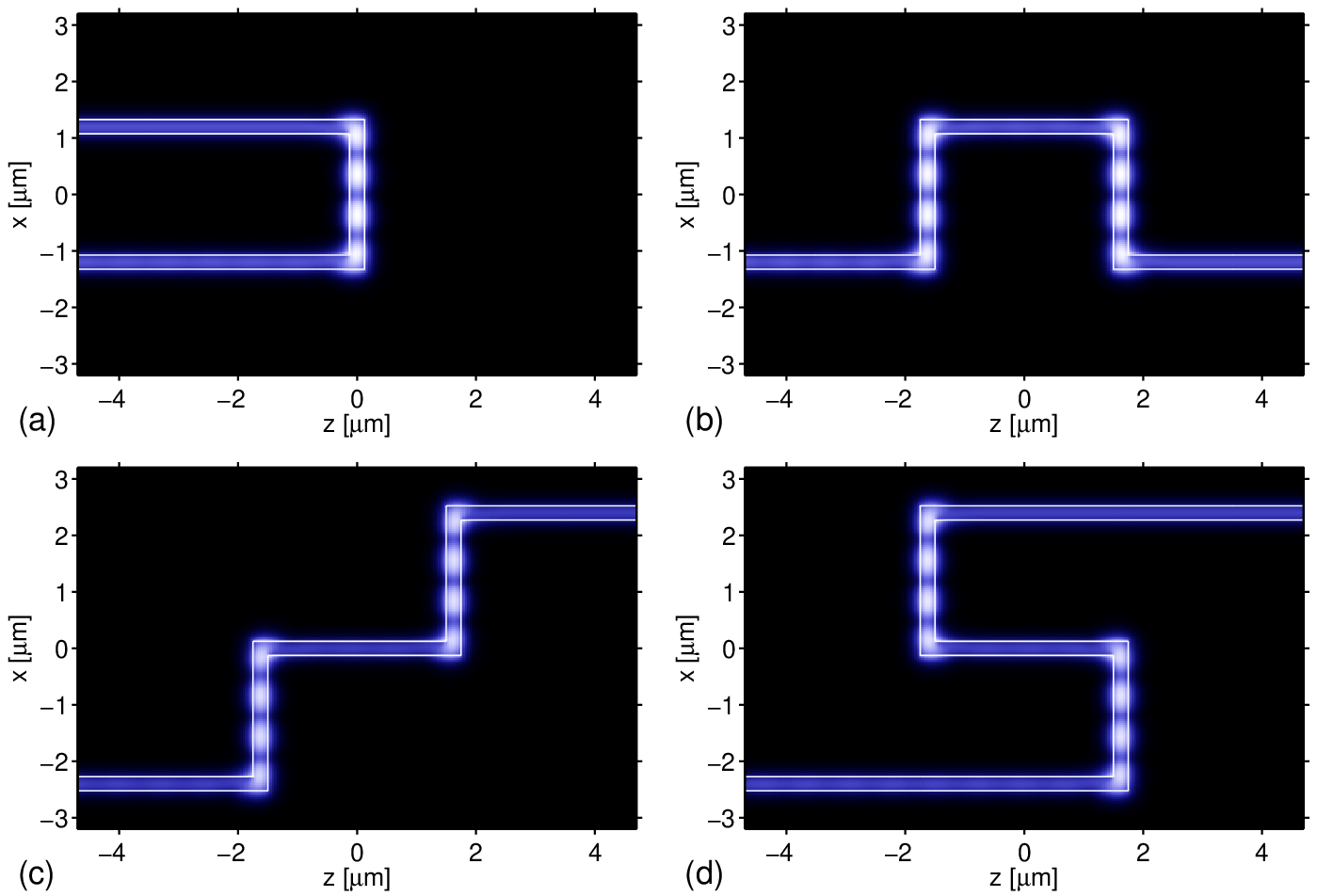}{12.8cm}{Field profiles $|\nv{E}|$ for u-turn, bridge, stair, and s-bend -like structures, each made of double-corners with 
parameters as for configuration (C1) of \refsec{dkink} \protect\\
($\theta = 68^\circ$,  $h = 2.15\mum$), with an (arbitrary) $z$-distance of $3\mum$
between the vertical sections in (b, c, d).
}

With the exception of the height scan of \refsec{dkink}, design of the step structures is straightforward, 
without the necessity of further tuning of parameters. Still there is ample room for optimization: Apart from 
variations of layer thicknesses, other shapes could be investigated for the refractive index 
profile of the corner regions. One might aim at a higher transmittance of the single corners, and
consequently at a wider angular range of high transmittance for the steps with higher transmission of narrower 
bundles, preferably for single polarization configurations such as the present (C1). 
One might also consider structures with true lateral guiding. This would require the preparation of wide
channels with weak lateral contrast (e.g.\ by shallow etching, indiffusion, direct laser writing, 
or other processes) along the path of the present beams.


\section*{Acknowledgments}
{\em
Financial support from the German Research Foundation (Deutsche Forschungsgemeinschaft DFG, projects HA\,7314/1-1, GRK\,1464, and TRR\,142)
is gratefully acknowledged. 
}


\begin{small}
\bibliographystyle{unsrt}
\bibliography{wgcalc}

\end{small}


\end{sloppypar}
\end{document}